\documentclass[prl,twocolumn]{revtex4}

\begin{document}

\title{Low- and high-frequency noise from coherent two-level systems}
\author{Alexander Shnirman$^1$, Gerd Sch\"on$^1$, Ivar Martin$^2$, and Yuriy Makhlin$^3$}
\affiliation{
$^1$ Institut f\"{u}r Theoretische
Festk\"{o}rperphysik Universit\"{a}t Karlsruhe, D-76128 Karlsruhe,
Germany\\
$^2$Theoretical Division, Los Alamos National
Laboratory, Los Alamos, NM 87545, USA\\
$^3$Landau Institute for Theoretical Physics, Kosygin st. 2,
119334 Moscow, Russia}

\date{\today}

\begin{abstract}
Recent experiments indicate a connection between the low- and
high-frequency noise affecting superconducting quantum systems. 
We explore the possibilities that both noises can be
produced by one ensemble of microscopic modes, made up, e.g., by
sufficiently coherent two-level systems (TLS). This implies a
relation between the noise power in different frequency domains,
which depends on the distribution of the parameters of the TLSs.
We show that a distribution, natural for tunneling TLSs, with a
log-uniform distribution in the tunnel splitting and linear
distribution in the bias, accounts for experimental observations.
\end{abstract}


\maketitle

Recent activities and progress with quantum information systems
rely on the control of decoherence processes and at the same time
provide novel tools to study their mechanisms. Experiments with
superconducting qubits revealed the presence of spurious quantum
two-level systems~\cite{Simmonds_PRL04} with strong effects on the
high-frequency ($\sim$10\ GHz) qubit dynamics. Other
experiments~\cite{Astafiev} suggested a connection between the
strengths of the Ohmic high-frequency noise, responsible for the
relaxation of the qubit ($T_1$ decay), and the low-frequency $1/f$
noise, which dominates the dephasing ($T_2$ decay). The noise
power spectra, extrapolated from the low- and high-frequency
sides, cross at $\omega$ of order $T$. This is also compatible
with the $T^2$ dependence of the low-frequency part, observed
earlier for the $1/f$ noise in Josephson
devices~\cite{Wellstood_Thesis,Wellstood_T2}. Much clearer evidence 
for the $T^2$ behavior was obtained 
recently~\cite{Astafiev_Private,Wellstood_Clarke}.

In this letter we point out that a set of {\sl coherent} two-level
systems (or, in fact, arbitrary quantum systems with discrete
spectrum) produces both high- and low-frequency noise with
strengths that are naturally related.  We show that for a
realistic distribution of parameters tunnel TLSs (TTLS) produce
noise with experimentally detected features: the $1/f$ behavior at
low frequencies, the Ohmic ($\propto\omega$) high-frequency noise,
and the $T^2$ temperature dependence of the integrated weight of
the low-frequency noise. This implies that the $1/f$ and Ohmic
asymptotes cross at $\omega\sim T$ as was indeed observed in
Ref.~\cite{Astafiev} at one value of $T$. The distribution is
log-uniform in the tunnel splitting and linear in the bias.
Microscopically, this distribution may describe double traps or
``Andreev fluctuators" considered recently by Faoro {\it et
al.}~\cite{Faoro123} in their study of the relaxation ($T_1$
decay) of Josephson qubits due to the high-frequency noise. Our
results are obtained for environments with a large number of TLSs
which are weakly coupled to the qubit. A strong coupling between a
TLS and a qubit can lead to
resonances~\cite{Simmonds_PRL04,Astafiev}.

Ensembles of TLSs were discussed extensively in the literature. On one hand,
they produce a natural model of $1/f$ noise, as a result of
{\it incoherent} random transitions~\cite{Dutta_Horn},
and there is substantial experimental evidence
that the low-frequency $1/f$ noise in single-electron devices may be produced by
TLSs~\cite{Martinis_Flicker,Zorin_Flicker}.
In solid-state qubits, e.g., Josephson qubits, the pure dephasing is dominated
by this noise~\cite{Nakamura_Echo,Saclay_Science,Clarke}. On the other hand,
ensembles of coherent TTLSs were suggested to explain low-temperature properties of
glasses~\cite{AHV_TLS,Phillips_TLS}. Both ``transverse'' and ``longitudinal''
couplings, defined below, were discussed in relation to various physical phenomena.
A transverse coupling of phonons or electrons to the TLSs is
responsible for the absorption and emission of energy.  It was
invoked in the discussions of, e.g., the phonon attenuation~\cite{Black_Review}
and of the low-temperature dephasing in
disordered metals~\cite{Imry}. On the other hand a longitudinal
coupling was found to be responsible, e.g., for the conductance
fluctuations~\cite{Kogan,Ludviksson,Feng}. We suggest that in nanocircuits,
e.g., solid-state qubits, both types of couplings play an important part
and produce noise with related properties in various frequency ranges.

As a model we consider a set of coherent two-level systems
described by the Pauli matrices $\sigma_{p,j}$, where $p=x,y,z$
and $j$ is the index of a particular TLS. We write the Hamiltonian of the set
in the basis defined by their contributions to the relevant fluctuating
quantity (cf.~Eq.~(\ref{Eq:X}) below);
\begin{equation}
H_{\rm TLS} = \sum_j\,\left[-\frac{1}{2}\,\left( \varepsilon_j \sigma_{z,j} +
\Delta_j \sigma_{x,j}\right) + H_{{\rm diss},j}\right] \ .
\end{equation}
Here, in the language of TTLSs, $\varepsilon_j$ are the bias energies and
$\Delta_j$ are the tunnel amplitudes between two states.
Each TLS with label $j$ is subject to dissipation due to its own
bath with Hamiltonian $H_{{\rm diss},j}$. We do not specify
$H_{{\rm diss},j}$, but only assume that it produces the usual
relaxation ($T_1$) and  dephasing ($T_2$) processes.
We assume that all the TLSs are under-damped, with
$\Gamma_{1,j}\equiv T_{1,j}^{-1}\ll E_j$ and
$\Gamma_{2,j}\equiv T_{2,j}^{-1} \ll E_j$. Here $E_j \equiv
\sqrt{\epsilon_j^2 + \Delta_j^2}$ is the energy splitting.

Each TLS in the ensemble contributes to fluctuations of a physical quantity
$X$, e.g., the gate charge, which affect an experimentally accessible system
and thus may be detected. A qubit may serve as a convenient noise
detector~\cite{Aguado,SchoelkopfNoiseDetector}. E.g., in the recent experiment of
Ref.~\cite{Astafiev} qubits were used to investigate the
properties of their environment. We choose
\begin{equation}
X\equiv \sum_j v_j\, \sigma_{z,j} \ ,
\label{Eq:X}
\end{equation}
where $v_j$ are the coupling constants and $\sigma_{z,j} = \pm 1$
correspond to the two states differing, e.g., by the value of the
dipole moment. The interaction of the qubit with the TLSs
is often described via a linear in $X$ coupling to a variable $O_{\rm qubit}$
of the qubit, i.e., $H_{int}(X)\propto X O_{\rm qubit}$.

Our goal in the following is to investigate the noise properties
of $X$, that is we need to evaluate the (unsymmetrized) correlator
\begin{equation}
C_X(\omega) \equiv \int dt\, \left\{\langle
X(t)X(0)\rangle - \langle X\rangle^2\right\}\, e^{i\omega t} \ .
\end{equation}
For independent TLSs the noise is a sum of
individual contributions, $C_X = \sum_j v_j^2\,C_j$, where
\begin{equation}
C_j(\omega) \equiv \int dt \left\{\langle
\sigma_{z,j}(t)\sigma_{z,j}(0)\rangle - \langle
\sigma_{z,j}\rangle^2 \right\} e^{i\omega t} \ .
\end{equation}
To obtain $C_j$ we first transform to the eigenbasis of the TLS.
This gives
\begin{equation}
H_{\rm TLS} =\sum_j\,\left\{ -\frac{1}{2}\, E_j \rho_{z,j} + H_{{\rm
diss},j}\right\} \ ,
\end{equation}
and
\begin{equation}
\label{eq:X}
X =\sum_j v_j\,(\cos\theta_j\,\rho_{z,j}-\sin\theta_j\,\rho_{x,j}) \ ,
\end{equation}
where $\tan\theta_j \equiv \Delta_j/\epsilon_j$.
The first term of (\ref{eq:X}) produces the longitudinal coupling (mentioned above)
while the second term produces the transverse one. Using the
Bloch-Redfield theory~\cite{Bloch_Derivation,Redfield_Derivation} we find
readily
\begin{eqnarray}
\label{eq:S_j} C_j(\omega) &\approx& \cos^2\theta_j\,\left[
1-\langle \rho_{z,j} \rangle^2\right]
\frac{2\Gamma_{1,j}}{\Gamma_{1,j}^2 + \omega^2}
\nonumber \\
&+&
\sin^2\theta_j\,\left[\frac{1+\langle\rho_{z,j}\rangle}{2}\right]\,\frac{2\Gamma_{2,j}}
{\Gamma_{2,j}^2 + (\omega-E_j)^2}
\nonumber \\
&+&
\sin^2\theta_j\,\left[\frac{1-\langle\rho_{z,j}\rangle}{2}\right]\,\frac{2\Gamma_{2,j}}
{\Gamma_{2,j}^2 + (\omega+E_j)^2} \ .
\end{eqnarray}
In thermal equilibrium $\langle\rho_{z,j}\rangle = \tanh(E_j/2T)$.
The first term, due to the longitudinal part of the coupling,
describes random telegraph noise of a thermally excited TLS. 
We have assumed $\Gamma_{1,j} \ll T$, so that this term is 
symmetric (classical).
The second term is due to the transverse coupling and describes
absorption by the TLS, while the third term describes the
transitions of the TLS with emission. We observe that TLSs with
$E_j \gg T$ contribute to $C_X$ only at (positive) $\omega = E_j$. Indeed
their contribution at $\omega=0$ is suppressed by the thermal
factor $1-\langle \rho_{z,j}\rangle^2 = 1-\tanh^2(E_j/2T)$. 
Also the negative frequency (emission) contribution at $\omega = - E_j$
is suppressed. These high-energy TLSs remain always in their ground state. Only the
TLSs with $E_j < T$ are thermally excited, performing real random
transitions between their two eigenstates, and contribute both at
$\omega = \pm E_j$ and at $\omega=0$. Such a multi-peaked
structure of $C_j(\omega)$ was recently discussed in various
contexts, e.g., in
Refs.~\cite{AverinKorotkov,ShnirmanMozyrskyMartin,GalperinPeaks}.
Note that the separation of the terms in Eq.~(\ref{eq:S_j}) into
low- and high-frequency noise is meaningful provided the typical
width $\Gamma_{1,j}$ of the low-$\omega$ Lorentzians is lower than
the high frequencies of interest, which are defined, e.g., by the
qubit's level splitting or temperature.

For a dense distribution of the parameters $\epsilon$, $\Delta$,
and $v$ we can evaluate the low- and high-frequency
noise. For positive high frequencies, $\omega \gg T$, we obtain
\begin{eqnarray}
\label{eq:CXomega_general}
&&C_X(\omega) \approx \sum_j\,v_j^2
\sin^2\theta_j\;\frac{2\Gamma_{2,j}} {\Gamma_{2,j}^2 +
(\omega-E_j)^2}\nonumber \\&&\approx
N\int d\epsilon d\Delta d v \,P(\epsilon,\Delta,v)\, v^2 \sin^2\theta \cdot
2\pi\delta(\omega - E)
\ ,\nonumber\\
\end{eqnarray}
where $N$ is the number of fluctuators, $P(\epsilon,\Delta,v)$
is the distribution function normalized to 1,
$E\equiv \sqrt{\epsilon^2 + \Delta^2}$,
and $\tan\theta = \Delta/\epsilon$.
Without loss of generality we take $\epsilon \ge 0$ and $\Delta \ge 0$.
At negative high frequencies ($\omega < 0$ and $|\omega| > T$) the 
correlator $C_X(\omega)$ is exponentially suppressed. 

On the other hand,
the total weight of the low-frequency (up to $\omega \gtrsim \Gamma_1$) noise follows from
the first term of (\ref{eq:S_j}). Each Lorentzian contributes
$1$. Thus we obtain
\begin{eqnarray}
\label{eq:intCXlf_general}
&&\int\limits_{\rm low\ freq.} \frac{d\omega}{2\pi}\; C_X(\omega)
\nonumber\\&&\approx \int\limits_{\rm low\ freq.}
\frac{d\omega}{2\pi}\; \sum_j\;v_j^2 \cos^2\theta_j\,\left[
1-\langle \rho_{z,j} \rangle^2\right]
\frac{2\Gamma_{1,j}}{\Gamma_{1,j}^2 + \omega^2}
\nonumber\\&&\approx N\int d\epsilon d\Delta d v \,P(\epsilon,\Delta,v)\, v^2
\cos^2\theta\;\frac{1}{\cosh^2\frac{E}{2T}}
\ .\nonumber\\
\end{eqnarray}
Here we could disregard the contribution of the last two terms for $E_j \sim \Gamma_{1,j}$.
Equations (\ref{eq:CXomega_general}) and (\ref{eq:intCXlf_general})
provide the general framework for further discussion.

Next we investigate possible distributions for the parameters $\epsilon$, $\Delta$, and $v$.
We consider a log-uniform distribution of tunnel splittings $\Delta$, with density
$P_\Delta(\Delta)\propto 1/\Delta$ in a range $[\Delta_{\rm min},\Delta_{\rm max}]$.
This distribution is natural for TTLSs as $\Delta$ is an exponential function
of an almost uniformly distributed parameter, e.g., tunnel barrier height~\cite{Phillips_TLS}.
It is also well known to provide for the $1/f$ behavior of the low-frequency
noise~\cite{Dutta_Horn}: the relaxation rates are, then, also
distributed log-uniformly, $P_{\Gamma_1}(\Gamma_1) \propto 1/\Gamma_1$, and a sum of many
Lorentzians of width $\Gamma_1$ centered at $\omega=0$ adds up to the $1/f$ noise.
We further assume that the distribution of $v$ is
uncorrelated with $\varepsilon$ and $\Delta$.

First, we assume that the temperature is lower than $\Delta_{\rm max}$,
$T<\Delta_{\rm max}$.
For the high-frequency part, $T < \omega < \Delta_{\rm max}$, taking the integral over
$\Delta$ in Eq.~(\ref{eq:CXomega_general}), we find that
\begin{equation}
C_X(\omega) \propto \frac{1}{\omega} \int_0^\omega P_\varepsilon(\varepsilon) d\varepsilon
\,.
\end{equation}
This is consistent with an Ohmic behavior $C_X \propto \omega$ only for the linear
density $P_\varepsilon(\varepsilon) \propto \varepsilon$.

Importantly, this distribution $P(\varepsilon, \Delta) \propto
\varepsilon / \Delta$ leads at the same time to the $T^2
\ln(T/\Delta_{min})$ behavior of the low-frequency weight
(\ref{eq:intCXlf_general}), consistent with experimental
observations~\cite{Wellstood_Thesis,Wellstood_T2,
Astafiev_Private,Wellstood_Clarke}. 
If the low-frequency noise has a $1/f$
dependence, the two parts of the spectrum would cross around
$\omega\sim T$~\cite{Astafiev}.

A remark is in order concerning this crossing.
It is not guaranteed that the spectrum has a $1/f$ dependence up to
$\omega \sim T$. Rather the high-frequency cutoff of the low-frequency
$1/f$ noise is given by the maximum
relaxation rate of the TLSs, $\Gamma_{1,{\rm max}}\ll T$, as we assumed.
Then the {\sl extrapolations} of the low-frequency $1/f$ and
high-frequency Ohmic spectra cross
at this $\omega\sim T$.

To fix the coefficients, we introduce the normalization constant $A$,
so that $P(\epsilon,\Delta) = A\epsilon/\Delta$.
Then, at high positive frequencies, $T < \omega < \Delta_{\rm max}$, we obtain
\begin{eqnarray}
\label{eq:CXomega}
C_X(\omega) \approx \pi \langle v^2 \rangle N A\, \omega
\ .
\end{eqnarray}
For the total weight of the low-frequency noise we obtain
\begin{eqnarray}
\label{eq:intCX}
\int\limits_{\rm low\ freq.} \frac{d\omega}{2\pi}\; C_X(\omega)
\approx 4\ln(2)\,\langle v^2 \rangle N A T^{2}\,
\left[\ln\frac{T_{\phantom{min}}}{\Delta_{\rm min}}\right]\ .
\end{eqnarray}
Thus we obtain a numerical factor which determines precisely
the point of crossing of the two spectra.

In the opposite limit, $T\gg \Delta_{\rm max}$, the high/frequency
noise depends on the detailed shape of the cutoff of $P_\Delta
(\Delta)$ at $\Delta_{\rm max}$. As an example, for a hard cutoff
the Ohmic spectral density implies that $P_\varepsilon \propto
\varepsilon^3$, and the low-frequency weight scales with $T^4$.
For a $1/f$ low-frequency behavior, the spectra would cross at
$\omega\sim T^2 / \Delta_{\rm max} \gg T$, in disagreement with
the result of Ref.~\cite{Astafiev}.

Interestingly, the linear $\omega$ dependence at high frequencies
and the $T^2$ dependence of the low-frequency noise can be
obtained from a whole class of distributions,
e.g., for $P(\epsilon,\Delta)=f(\epsilon/\Delta)$, with arbitrary, not too
divergent (as a function of $\theta$) function $f$.
Presented as a function of energy $E$ and angle $\theta$, it
becomes (we have used the Jacobian $d\epsilon d\Delta \rightarrow E dE d\theta$)
\begin{equation}
P(E,\theta) = E\,f(\cot\theta)
\ ,
\end{equation}
i.e., it is linear in $E$. This linearity ensures both the linear
$\omega$ dependence at high frequencies and the $T^2$ dependence
of the integrated weight of the low-frequency noise.

In particular one can take $P(\epsilon,\Delta) \propto
(\epsilon/\Delta)^s$ with any exponent $s$ satisfying $-1\le s \le 1$, and
with both $\Delta_{\rm max}$ and $\epsilon_{\rm max}$ higher
than the relevant frequency $\omega$. This includes a
uniform distribution of both $\Delta$ and $\epsilon$ at all
relevant energies, i.e., $s=0$.
For ensembles with $-1<s<1$, including
the uniform distribution with $s=0$,
high- and low-frequency noise is created by the same
fluctuators. On the other hand, $s=1$ is the limiting case in
which the low-frequency noise is dominated by the fluctuators with
$\theta \ll 1$, while the high-frequency noise by all other
fluctuators. Yet, even in this case the strengths of the high-
and low-frequency parts of the spectrum are related.

For the uniform distribution ($s=0$) we obtain different
numerical coefficients. We introduce an experimentally accessible
constant $\alpha$, such that $C_X(\omega \gg T) = 2\pi\alpha\omega$. Then,
for $s=1$ ($P\propto \epsilon/\Delta$) we obtain from
Eqs.~(\ref{eq:CXomega}) and (\ref{eq:intCX}) that
the total weight of the low-frequency noise is given by $\int_{\rm
l.f.} \frac{d\omega}{2\pi}\; C_X(\omega) \approx 8\ln(2)\alpha T^2
\ln(T/\Delta_{\rm min})$. On the other hand, for the uniform distribution,
$s=0$, we obtain $\int_{\rm l.f.} \frac{d\omega}{2\pi}\; C_X(\omega)
\approx 4\ln(2)\alpha T^2$.

We would like to emphasize that the relation between low- and
high-frequency noise is more general, i.e., it is not unique to
an ensemble of two-level systems. Consider an ensemble of
many-level systems with levels $|n\rangle$ and energies $E_n$
such that the coupling is via an observable
which has both transverse and longitudinal components. By a
transverse component we mean the part constructed with operators
$|n\rangle \langle m|$, where $n\neq m$, while the longitudinal
component is built from the projectors $|n\rangle \langle n|$.
If the system is under-damped, that is, if the absorption and
emission lines are well defined, the correlator of such an
observable will have (Lorentzian-like) contributions at $\omega =
E_n-E_m$ as well as at $\omega=0$. As an example we could consider
an ensemble of an-harmonic oscillators
with $X=\sum_{j} v_j x_j$, where $x_j$ are the oscillator's
coordinates. Due to the anharmonicity $x_j$ acquires a
longitudinal component, in addition to the usual transverse one.
Thus a relation between the low- and high-frequency noise would
emerge naturally with details depending on the ensemble statistics.

It is useful to relate our phenomenological results to the recent
work of Faoro {\it et al.}~\cite{Faoro123}, where they considered
physical models of the fluctuators, which could couple to and
relax qubits.  They considered three models: (I) a single electron
trap in tunnel contact with a metallic gate, (II) a single
electron occupying a double trap, and (III) a double trap that can
absorb/emit a Cooper pair from the qubit or a superconducting
gate. In all models a uniform distribution of the trap energy
levels was assumed. One, then, can show that the distribution for
the  two-level systems corresponding to the models II and III are
linear in the energy level splitting, $P(\epsilon) \propto
\epsilon$.  Since the switching in these models is tunneling
dominated, we find that $P(\Delta) \propto 1/\Delta$.  Therefore,
both models II and III are characterized by distribution
$P(\epsilon,\Delta)=A\epsilon/\Delta$, described above, and hence
can naturally account for the experimentally observed low- and
high-frequency noises.  In contrast, in the model of uniformly
distributed single-electron traps (model I), we find that for
small tunnel rates, the high-frequency noise is inversely
proportional to frequency rather than Ohmic~\cite{Faoro123}.

In this letter we did not address the question of the statistics
of the low-frequency noise, nor the associated problem of a
particular decay law of the dephasing process. These
statistics will depend on the distribution of the coupling
strengths $v_j$. For certain distributions the individual strongly
coupled fluctuators may be important~\cite{Paladino_PRL02,Galperin},
and the statistics is non-Gaussian. For ensembles of many weakly
coupled fluctuators Gaussian statistics
emerges~\cite{Nakamura_Echo,Our_PhysScripta}.

To conclude, we have shown that an ensemble of coherent two-level systems
with the distribution function, $P(\epsilon,\Delta) \propto \epsilon/\Delta$,
produces Ohmic high-frequency noise and, at the same time,
$1/f$ low-frequency noise with strength which grows with temperature as $T^2$.
The two branches of
the noise power cross at $\omega\sim T$ in
accordance with the experimental observation~\cite{Astafiev}.
A relation between low- and high-frequency parts of the spectrum  is a general
property of ensembles of coherent systems with discrete energy levels.

We are grateful to Yu.~A.~Pashkin for communicating the
experimental results of Ref.~\cite{Astafiev} prior to publication.
We thank J. Clarke and O. Astafiev for stimulating discussions and 
useful comments. The work is part of the CFN of the DFG and of the
EU IST Project SQUBIT. This work was supported by the U.S. DoE.
YM acknowledges support from the Dynasty Foundation.

\bibliographystyle{apsrev}
\bibliography{ref}

\end{document}